\documentclass{article}

\normalbaselines

\usepackage{latexsym}

\newtheorem{theorem}{Theorem}[section]
\newtheorem{define}[theorem]{Definition}
\newenvironment{definition}{\begin{define} \rm}{\end{define}}
\newtheorem{exa}[theorem]{Example}

\newtheorem{lemma}[theorem]{Lemma}
\newtheorem{corollary}[theorem]{Corollary}

\newtheorem{exe}{Exercise}

\newcommand{\diag}[2]{d_{{#1}{#2}}}

\newcommand{\rarrow}{\rightarrow}

\newcommand{\la}{\langle}
\newcommand{\ra}{\rangle}
\newcommand{\lan}{\langle}
\newcommand{\ran}{\rangle}
\newcommand{\os}{[\![}
\newcommand{\cs}{]\!]}
\newcommand{\rrarrow}{\longrightarrow}

\newcommand{\tell}{{\bf tell}}

\newcommand{\ask}{{\bf ask}}

\newcommand{\sat}{{\it sat}}

\newcommand{\true}{{\it true}}
\newcommand{\false}{{\it false}}
\newcommand{\vars}{{\it Var}}

\long\def\comment#1{}

\begin{document}

\title{Proving correctness of Timed Concurrent Constraint Programs}

\author{
F.S. de Boer
\\
Universiteit Utrecht
\thanks{Department of Computer Science,
Universiteit Utrecht, Padualaan 14, De Uithof, 3584 CH Utrecht, The Netherlands. {\tt frankb@cs.uu.nl}}
\addtocounter{footnote}{2}
        \and
M. Gabbrielli
\\
Universit\`{a} di Bologna
\thanks{Dipartimento di Scienze dell'Informazione,
Mura A. Zamboni 7, 40127 Bologna, Italy. {\tt gabbri@cs.unibo.it.} }
        \and
M.C. Meo
\\
Universit\`{a} di Chieti
\thanks{Dipartimento di Scienze,
Universit\`{a} di Chieti, Viale Pindaro 42, Pescara, 65127, Italy.
{\tt meo@unich.it.}} }

\date{}

\maketitle

\begin{abstract}
\noindent A temporal logic is presented for reasoning about the
correctness of timed concurrent constraint programs. The logic is
based on modalities which allow one to specify what a process
produces as a reaction to what its environment inputs.
These modalities provide
an assumption/commitment style of specification which allows
a sound and complete
compositional axiomatization of the reactive behavior of timed
concurrent constraint programs.

\end{abstract}

\section{Introduction}\label{sec:intro}

Many ``real-life'' computer applications maintain  some ongoing
interaction with external physical processes and involve
time-critical aspects. Characteristic
of such applications, usually called
real-time embedded systems, is the specification of  timing
constraints such as, for example, that an input is required
within a bounded period of time.
Typical examples of such systems are
process controllers and signal processing systems.


In \cite{BGM00} {\it tccp}, a timed extension of the pure
formalism of concurrent constraint programming(\cite{Sa89a}), is
introduced. This extension is based on the hypothesis of {\em
bounded asynchrony} (as introduced in \cite{SJG96}): Computation
takes a bounded period of time rather than being instantaneous
as in the concurrent synchronous languages {\tt ESTEREL}
\cite{BG92}, {\tt LUSTRE} \cite{HCP91}, {\tt SIGNAL}
\cite{GBGM91} and Statecharts \cite{Ha87}. Time itself is
measured by a discrete global clock, i.e, the internal clock of
the {\it tccp} process. In \cite{BGM00} we also introduced {\em
timed reactive sequences} which describe at each moment in time
the reaction of a tccp process to the input of the external
environment. Formally, such a reaction is a pair of constraints
$\langle c,d\rangle$, where $c$ is the input given by the
environment and $d$ is the constraint produced by the process in
response to the input $c$ (such a response includes always the
input because of the monotonicity of ccp computations).

In this paper we introduce a  temporal logic for describing
and reasoning about
timed reactive sequences.
The basic assertions of the temporal logic
describe the reactions of such a sequence in terms of  {\em modalities}
which express either what a process {\em assumes}
about the inputs of the environment and what a process
{\em commits} to, i.e., has itself produced at
one time-instant. These modalities
thus provide a kind of
assumption/commitment style of specification of the reactive
behavior of a process. The main result of
this paper is a sound and complete
compositional proof system for reasoning about
the correctness of {\it tccp} programs as specified by formulas
in this temporal logic.

The remainder of this paper is organized as follows. In the next section we introduce the language {\it tccp} and
its operational semantics. In Section 3 we introduce the temporal logic and the compositional proof system. In
Section 4 we discuss soundness and completeness of the proof system. Section 5 concludes by discussing related
work and indicating future research. A preliminary, short version of this paper appeared in \cite{BGM02}.

\section{The programming language}\label{sec:language}

In this section we first define the {\it tccp} language and then
we define formally its operational semantics  by using a
transition system.

Since the starting point is ccp, we introduce first some basic
notions related to this programming paradigm. We refer to
\cite{SR90,SRP91} for more details. The ccp languages are
defined parametrically wrt to a given {\em constraint system}.
The notion of constraint system has been formalized in
\cite{SR90} following Scott's treatment of information systems.
Here we only consider the resulting structure.

\begin{definition}
A constraint system is a complete algebraic lattice
$\lan {\cal C}, \leq, \sqcup,\true,\false\ran$ where
$\sqcup$ is the lub operation, and  $\true$,  $\false$
are the least and the greatest elements of ${\cal C}$, respectively.
\end{definition}

Following the standard terminology and notation,
instead of $\leq$ we will refer to its inverse relation,
denoted by $\vdash$ and called {\it entailment}. Formally, $\forall c,d\in C.\;
\;
c\vdash d
\;\Leftrightarrow\;
d\leq c.$
In order to treat the hiding operator of the language
a general notion of existential quantifier
is introduced which is formalized in terms of
cylindric algebras \cite{HMT71}. Moreover,
in order to model parameter passing,
{\it diagonal elements} \cite{HMT71} are added
to the primitive constraints. This leads to
the concept of
a {\it cylindric constraint system}. In the following, we assume given a
(denumerable) set
of variables ${\it Var}$  with typical elements $x,y,z,\ldots$.

\begin{definition}\label{Ccs}
Let $\lan {\cal C}, \leq, \sqcup,\true,\false \ran$ be a constraint
system. Assume that for each $x\in \vars$  a function $\exists_x:{\cal C}
\rarrow
{\cal C}$ is defined such that for any $c,d \in {\cal C}$:
$$
\begin{array}[t]{ll}
{\rm (i)}\ c\vdash \exists_x (c), &{\rm(ii)} \hbox{
if \ $c\vdash d$ \ then \ $\exists_x (c)\vdash\exists_x (d) $},
\\
{\rm(iii)}\ \exists_x(c\sqcup \exists_x(d)) = \exists_x(c)\sqcup \exists_x(d),
&{\rm(iv)} \ \exists_x(\exists_y(c))=\exists_y(\exists_x(c)).
\end{array}
$$
Moreover assume
that for $x,y$ ranging in $\vars$, ${\cal C}$ contains the
constraints $d_{xy}$ (so called diagonal elements)
which satisfy the following axioms:
$$
\begin{array}[t]{ll}
{\rm (v)} \ true \vdash d_{xx},
& {\rm (vi)}
\hbox{ if $z\neq x,y$ then $d_{xy} = \exists_z(d_{xz}\sqcup d_{zy})$},
\\
{\rm (vii)}
\hbox{ if $x\neq y$ then $d_{xy}\sqcup\exists_x (c\sqcup d_{xy})\vdash c $}.
\end{array}
$$
Then {\bf C} =
$\lan {\cal C}, \leq, \sqcup,\true,\false ,\vars,\exists_x,d_{xy}\ran$ is a
{\it cylindric constraint system}.
\end{definition}

Note that if ${\bf C}$ models the equality theory, then the
elements $d_{xy}$ can be thought of as the formulas $x=y$. In
the sequel we will identify a system {\bf C} with its underlying
set of constraints ${\cal C}$ and we will denote  $\exists_x(c)$
by $\exists_x c$ with the convention that, in case of ambiguity,
the scope of $\exists_x$ is limited to the first constraint
sub-expression (so, for instance, $\exists_x c\sqcup d$ stands
for $\exists_x(c)\sqcup d$).

The basic idea underlying ccp is that computation
progresses via monotonic accumulation of information
in a global store. Information is produced by the concurrent and asynchronous
activity of several agents which can add ({\em tell}) a constraint to the
store.
Dually, agents can also
check ({\it ask}) whether a constraint is entailed by the store, thus
allowing synchronization among different agents.
Parallel composition in ccp is modeled by the interleaving
of the basic actions of its components.

When querying the store for some information which is not
present (yet) a ccp  agent will simply suspend until the
required information has arrived. In timed applications however
often one cannot  wait indefinitely for an event. Consider for
example the case of a bank teller machine. Once a card is
accepted and its identification number has been checked, the
machine asks the authorization of the bank to release the
requested money. If the authorization does not arrive within a
reasonable amount of time, then the card should be given back to
the customer. A timed language should then allow us to specify
that, in case a given time bound is exceeded (i.e. a {\em
time-out} occurs), the wait is interrupted and an alternative
action is taken. Moreover in some cases it is also necessary to
abort an active process $A$ and to start a process $B$ when a
specific event occurs (this is usually called {\em preemption}
of $A$). For example, according to a typical pattern, $A$ is the
process controlling the normal activity of some physical device,
the event indicates some abnormal situation and $B$ is the
exception handler.

In order to be able to specify these timing constraints in ccp
we introduce a discrete global clock and assume that
$ask$ and $tell$ actions take one time-unit.
Computation evolves in steps of one time-unit,
so called clock-cycles. We consider action prefixing as the
syntactic marker which distinguishes
a time instant from the next one. Furthermore we make the assumption
that parallel processes are executed on different processors,
which implies that at each moment every enabled agent
of the system is activated.
This assumption gives rise to what is called
{\em maximal parallelism}. The time in between two successive moments of the
global clock
intuitively corresponds to the response time of the
underlying constraint system.
Thus essentially in our model all parallel agents are
synchronized by the response time of the
underlying constraint system.

Furthermore, on the basis of the above assumptions we introduce a
timing construct of the form {\bf now} $c$ {\bf then} $A$ {\bf
else} $B$ which can be interpreted as follows: If the constraint
$c$ is entailed by the store at the current time $t$ then the
above agent behaves as $A$ at time $t$, otherwise it behaves as
$B$ at time $t$. As shown in \cite{BGM00,SJG96} this basic
construct allows one to derive such timing mechanisms as time-out
and preemption. Thus we end up with the following syntax of timed
concurrent constraint programming.

\begin{definition}[{\it tccp} Language \cite{BGM00}]
Assuming a given cylindric constraint system {\bf C}
the syntax of
{\em agents} is given by the following grammar:
\[
\begin{array}{l}
A ::= \hbox{\tell}(c)\;|\;\sum_{i=1}^n \hbox{\bf ask}(c_i) \rarrow
A_i \;|\; {\bf now } \ c \ {\bf then  }\ A \ {\bf else } \ B \;|\;
A\parallel B \;|\;  \exists x A \;|\; p(x)
\end{array}
\]
where the $c,c_i$ are supposed to be {\em finite constraints}
(i.e. algebraic elements) in ${\cal C}$.
A {\em tccp} {\em process} $P$ is then an object of the form $D.A$, where
$D$ is a set of procedure declarations of the form $p(x):: A$ and
$A$ is an agent.
\end{definition}

Action prefixing is denoted by $\rightarrow$, non-determinism is
introduced via the guarded choice construct $\sum_{i=1}^n {\bf
ask}(c_i)\rarrow A_i$, parallel composition is denoted by
$\parallel$, and a notion  of locality is introduced by the agent
$\exists x A$ which behaves like $A$ with $x$ considered local to
$A$, thus hiding the information on $x$ provided by the external
environment. In the next subsection we describe formally the
operational semantics of {\it tccp}. In order to simplify the
notation, in the following we will omit the $\sum_{i=1}^n$
whenever $n=1$ and we will use $\hbox{\tell}(c)\rightarrow$ {\it
A} as a shorthand for $\hbox{\tell}(c) \parallel (\ask(true)
\rightarrow  A)$. In the following we also assume guarded
recursion, that is we assume that each procedure call is in the
scope of an $\ask$ construct. This assumption, which does not
limit the expressive power of the language, is needed to ensure a
proper definition of the operational semantics.

\subsection{Operational semantics}\label{sec:op}

The operational model of {\it tccp} can be formally described by a transition system $T= ({\it Conf}, \rrarrow )$
where we assume that each transition step takes exactly one time-unit. Configurations (in) {\it Conf} are pairs
consisting of a process and a constraint in ${\cal C}$ representing the common {\em store}. The transition
relation $\rrarrow \subseteq  {\it Conf} \times {\it Conf}$ is the least relation satisfying the rules {\bf
R1-R10} in Table \ref{t1} and characterizes the (temporal) evolution of the system. So, $\langle A,c\rangle
\rrarrow \langle B,d\rangle $ means that if at time $t$ we have the process $A$ and the store $c$ then at time
$t+1$ we have the process $B$ and the store $d$.

\begin{table*}[th]
\begin{center}
\begin{tabular}{|lll|}  \hline
&\mbox{   }&\mbox{   }
\\
\mbox{\bf R1}& ${\la \hbox{\tell}(c),d \ra\rrarrow\la {\bf
stop},c\sqcup d\ra}$ &
\\
&\mbox{   }&\mbox{   }
\\
\mbox{\bf R2}&
${\la \sum_{i=1}^n \ask(c_i)\rightarrow A_i
,d \ra\rrarrow\la A_j,d \ra}$
& $j\in [1,n]\;\hbox{\it and} \; d \vdash c_j$
\\
&\mbox{   }&\mbox{   }
\\
\mbox{\bf R3 }&
$\frac
{\displaystyle \la A,d \ra
\rrarrow
\la A',  d' \ra }
{\displaystyle
\begin{array}{l}
\la {\bf now} \ c \ {\bf then} \ A\ {\bf else} \ B,d\ra\rrarrow
\la A',  d' \ra
\end{array}}$  & $ d\vdash c$
\\
&\mbox{   }&\mbox{   }
\\
\mbox{\bf R4 }&
$\frac
{\displaystyle \la A,d \ra
\not \rrarrow }
{\displaystyle
\begin{array}{l}
\la {\bf now} \ c \ {\bf then} \ A\ {\bf else} \ B,d\ra\rrarrow
\la A,  d \ra
\end{array}}$  & $ d\vdash c$
\\
&\mbox{   }&\mbox{   }
\\
\mbox{\bf R5 }&
$\frac
{\displaystyle \la B,d\ra
\rrarrow
\la B',  d'  \ra }
{\displaystyle
\begin{array}{l}
\la {\bf now} \ c \ {\bf then} \ A\ {\bf else} \ B,d\ra\rrarrow
\la B',  d'\ra
\end{array}}$& $d\not\vdash c$
\\
&\mbox{   }&\mbox{   }
\\
\mbox{\bf R6 }&
$\frac
{\displaystyle \la B,d\ra
\not\rrarrow
 }
{\displaystyle
\begin{array}{l}
\la {\bf now} \ c \ {\bf then} \ A\ {\bf else} \ B,d\ra\rrarrow
\la B,  d\ra
\end{array}}$& $d\not\vdash c$
\\
&\mbox{   }&\mbox{   }
\\
\mbox{\bf R7 }&
$\frac
{\displaystyle \la A,c\ra
\rrarrow
\la A',  c' \ra\ \ \ \ \la B,c\ra
\rrarrow
\la B',  d' \ra }
{\displaystyle
\begin{array}{l}
\la A\parallel B,c\ra\rrarrow
\la A'\parallel B',  c'\sqcup d'\ra
\end{array}}$&\mbox{   }
\\
&\mbox{   }&\mbox{   }
\\
\mbox{\bf R8 }&
$\frac
{\displaystyle \la A,c\ra
\rrarrow
\la A',  c' \ra\ \ \ \ \la B,c\ra
\not\rrarrow}
{\displaystyle
\begin{array}{l}
\la A\parallel B,c\ra\rrarrow
\la A'\parallel B,  c'\ra
\\
\la B\parallel A,c\ra\rrarrow
\la B\parallel A',  c'\ra
\end{array}}$&\mbox{   }
\\
&\mbox{   }&\mbox{   }
\\
\mbox{\bf R9}&
$\frac
{\displaystyle \la A,d\sqcup\exists_x c\ra\rrarrow\la B, d' \ra}
{\displaystyle\la \exists^d x A,c\ra\rrarrow\la \exists^{d'} x B,
c\sqcup \exists_x d' \ra}$
&\mbox{   }\\
&\mbox{   }&\mbox{   }
\\
\mbox{\bf R10}&
$\frac
{\displaystyle \la A,c\ra\rrarrow\la B, d \ra}
{\displaystyle\la p(x),c\ra\rrarrow\la B, d\ra}$
&${\it  p(x)::A \in D}$\\
&\mbox{   }&\mbox{   }
\\
\hline
\end{tabular}
\caption{The transition system for ${\it tccp}$.}\label{t1}
\end{center}
\end{table*}

Let us now briefly discuss the rules in Table \ref{t1}. In order
to represent successful termination we introduce the auxiliary
agent ${\bf stop}$: it cannot make any transition. Rule {\bf R1}
shows that we are considering here the so called ``eventual''
tell: The agent $\hbox{\tell}(c)$ adds $c$ to the store $d$
without checking for consistency of $c\sqcup d$ and then stops.
Note that the updated store $c\sqcup d$ will be visible only
starting from the next time instant since each transition step
involves exactly one time-unit. According to rule ${\bf R2}$ the
guarded choice operator gives rise to global non-determinism: The
external environment can affect the choice since $\ask(c_j)$ is
enabled at time $t$ (and $A_j$ is started at time $t+1$) iff the
store $d$ entails $c_j$, and $d$ can be modified by other agents.
The rules ${\bf R3}$-${\bf R6}$ show that the agent {\bf now $c$
then $A$ else $B$} behaves as $A$ or $B$ depending on the fact
that $c$ is or is not entailed by the store. Differently from the
case of the ask, here the evaluation of the guard is
instantaneous: If $\langle A, d\rangle$ ($\langle B, d\rangle$)
can make a transition at time $t$ and $c$ is (is not) entailed by
the store $d$, then the agent {\bf now $c$ then $A$ else $B$} can
make the same transition at time $t$. Moreover, observe that  in
any case the control is passed either to $A$ (if $c$ is entailed
by the current store $d$) or to $B$ (in case $d$ does not entail
$c$). Rules {\bf R7} and {\bf R8} model the parallel composition
operator in terms of {\em maximal parallelism}: The agent
$A\parallel B$ executes in one time-unit all the initial enabled
actions of $A$ and $B$. Thus, for example, the agent $A:\
(\ask(c)\rightarrow {\bf stop})\parallel
(\hbox{\tell}(c)\rightarrow {\bf stop})$ evaluated in the store
$c$ will (successfully) terminate in one time-unit, while the same
agent in the empty store will take two time-units to terminate.
The agent $\exists x A$ behaves like $A$, with $x$ considered {\it
local} to $A$, i.e. the information on $x$ provided by the
external environment is hidden to $A$ and, conversely, the
information on $x$ produced locally by $A$ is hidden to the
external world. To describe locality in rule {\bf R9} the syntax
has been extended by an agent $\exists^d x A$ where $d$ is a local
store of $A$ containing information on $x$ which is hidden in the
external store. Initially the local store is empty, i.e. $\exists
x A=\exists^\true x A$.

Rule ${\bf R10}$ treats the case of a procedure call when the
actual parameter equals the formal parameter. We do not need more
rules since, for the sake of simplicity, here and in the following
we assume that the set {\it D} of procedure declarations is closed
wrt parameter names: That is, for every procedure call {\it p(y)}
appearing in a process {\it D.A} we assume that if the original
declaration for {\it p} in {\it D} is  {\it  p(x):: A} then {\it
D} contains also the declaration ${\it p(y)::} {\it \exists x
(\hbox{\tell}(\diag{x}{y}) \parallel A)}$\footnote{Here the
(original) formal parameter is identified as a local alias of the
actual parameter. Alternatively, we could have introduced a new
rule treating explicitly this case, as it was in the original ccp
papers.}.

Using the transition system described by (the rules in) Table
\ref{t1} we can now define our notion of observables which
associates with an agent a set of timed reactive sequences of the
form
$$\langle c_1,d_1\rangle \cdots \langle c_n,d_n\rangle \langle
d,d\rangle
$$
where a pair of constraints $\langle c_i,d_i\rangle$ represents
a reaction of the given agent at time i: Intuitively, the agent
transforms the global store from $c_i$ to $d_i$ or, in other
words, $c_i$ is the assumption on the external environment while
$d_i$ is the contribution of the agent itself (which includes
always the assumption). The last pair denotes a ``stuttering
step'' in which no further information can be produced by the
agent, thus indicating that a ``resting point'' has been reached.

Since the basic actions
of  {\it tccp} are monotonic and we can also model a new input
of the external environment by a corresponding tell operation, it is natural
to assume that reactive sequences are monotonically increasing.
So in the
following we will assume that each timed reactive sequence $\langle
c_1,d_1\rangle \cdots \langle c_{n-1},d_{n-1}\rangle\langle c_n,c_n\rangle$
satisfies the following condition: $d_i \vdash c_i \hbox{ and } \ c_j
\vdash d_{j-1},  $ for any $i\in [1,n-1]$ and $j\in [2,n]$.
Since the constraints arising from the reactions are finite, we also
assume that a reactive sequence contains only finite
constraints\footnote{Note that here we implicitly assume that if $c$ is a
finite element then also $\exists_x c$ is finite.}.

The set of all reactive sequences is denoted by ${\cal S}$ and its
typical elements by $s,s_1\ldots$, while sets of reactive sequences are denoted by $S,S_1\ldots$
and $\varepsilon$ indicates the
empty reactive sequence.
Furthermore, $\cdot$ denotes the operator which
concatenates sequences. Operationally the reactive sequences of an agent are generated as follows.
\begin{definition}\label{def:ra}
We define the semantics $R\in{\it Agent}\rightarrow{\cal P}({\cal S})$ by
\[
\begin{array}[t]{ll}
R({\it A})= & \{\la c,d\ra \cdot w \in {\cal S} \mid
\la {\it A}, c \ra \rightarrow \la {\it B}, d \ra
\hbox{ and } w\in
R(\it B) \}
\\
& \cup
\\
& \{\la c,c\ra \cdot w \in {\cal S} \mid \la {\it A}, c \ra \not \rightarrow
\hbox{ and } w\in R(A) \cup \{\varepsilon\}\}.
\end{array}
\]
\end{definition}

Note that $R(A)$ is defined as the union of the set of all
reactive sequences which start with a reaction of $A$ and the set
of all reactive sequences which start with a stuttering step of
$A$.  In fact, when an agent is blocked, i.e., it cannot react to
the input of the environment, a stuttering step is generated.
After such a stuttering step the computation can either continue
with the further evaluation of ${\it A}$ (possibly generating more
stuttering steps) or it can terminate, as a ``resting point'' has
been reached. These two case are reflected in the second part of
the definition of ${\it R(A)}$ by the two conditions $w\in R(A)$
and $w\in \{\varepsilon\}$, respectively. Note also that, since
the {\bf stop} agent used in the transition system cannot make any
move, an arbitrary (finite) sequence of stuttering steps is always
appended to each reactive sequence.

Formally $R$ is defined as the least fixed-point
of the corresponding operator $\Phi\in
({\it Agent}\rightarrow{\cal P}({\cal S}))
\rightarrow {\it Agent}\rightarrow{\cal P}({\cal S})$ defined by
\[
\begin{array}[t]{ll}
\Phi(I)({\it  A})= & \{\la c,d\ra \cdot w \in {\cal S} \mid
\la {\it A}, c \ra \rightarrow \la {\it B}, d \ra
\hbox{ and } w\in
I(\it B) \}
\\
& \cup
\\
& \{ \la c,c\ra \cdot w \in {\cal S} \mid \la {\it A}, c \ra \not \rightarrow
\hbox{ and } w\in I(\it A) \cup \{\varepsilon\}\}.
\end{array}
\]
The ordering on ${\it Agent}\rightarrow{\cal P}({\cal S})$
is that of (point-wise extended)
set-inclusion (it is straightforward to check that
$\Phi$ is continuous).

\section{A calculus for tccp}

In this section we introduce a temporal logic
for reasoning about the reactive behavior of
{\it tccp} programs. We first define
temporal formulas and the related notions of truth and validity
in terms of timed reactive sequences. Then we introduce the
correctness assertions that we consider and a corresponding proof system.

\subsection{Temporal logic}
Given a set $M$, with typical element $X,Y,\ldots$, of monadic
constraint predicate variables, our temporal logic is based on
atomic formulas of the form $X(c)$, where $c$ is a constraint of
the  given underlying constraint system. The distinguished
predicate $I$ will be used to express the ``assumptions'' of a
process about its inputs, that is, $I(c)$ holds if the process
assumes the information represented by $c$ is produced by its
environment. On the other hand, the distinguished predicate $O$
represents the output of a process, that is, $O(c)$ holds if the
information represented by $c$ is produced by the process itself
(recall that the produced information includes always the input,
as previously mentioned). More precisely, these formulas $I(c)$
and $O(c)$ will be interpreted with respect to a {\em reaction}
which consists of a pair of constraints $\la c,d\ra$, where $c$
represents the input of the external environment and $d$ is the
contribution of the process itself (as a reaction to the input
$c$) which always contains $c$ (i.e. such that $d\geq c$ holds).

An atomic formula in our temporal logic is a
formula as described above  or an atomic formula
of the form $c\leq d$ which
`imports' information about
the underlying constraint system, i.e., $c\leq d$ holds
if $d\vdash c$.
Compound formulas are constructed from these atomic formulas
by using the (usual) logical operators of
negation, conjunction and (existential) quantification and the
temporal operators $\bigcirc$ (the next operator) and ${\cal U}$
(the until operator).
We have the following three different kinds of quantification:
\begin{itemize}
\item
quantification over the variables $x,y,\ldots$
of the underlying constraint system;
\item
quantification over the constraints $c,d,\ldots$ themselves;
\item
quantification over the monadic constraint predicate variables $X,Y,\ldots$.
\end{itemize}
Variables $p,q,\ldots$ will range over the constraints.
We will use $V,W,\ldots$,
to denote a variable $x$ of the underlying constraint system,
a constraint variable $p$ or a constraint predicate $X$.

\begin{definition}[Temporal formulas]
Given an underlying constraint system with set of constraints
${\cal C}$,
formulas of the temporal logic are defined by
\[
\phi::= p\leq q\:\mid\;X(c)\;\mid\;\neg \phi\;\mid\;
\phi\wedge\psi\;\mid\;\exists V\phi\:|\; \bigcirc \phi\;\mid\;
\phi\;{\cal U}\; \psi
\]
\end{definition}

In the sequel we assume that the temporal operators have binding
priority over the propositional connectives. We introduce the
following abbreviations: $c=d$ stands for $c\leq d\wedge d\leq c$,
$\Diamond\phi$ for ${\it true}\;{\cal U}\;\phi$ and $\Box\phi$ for
$\neg\Diamond\neg\phi$. We also use $\phi \vee \psi$ as a
shorthand for $\neg (\neg \phi \wedge \neg \psi)$ and $\phi
\rightarrow \psi$ as a shorthand for $\neg \phi \vee \psi$.
Finally, given a temporal formula $\phi$, we denote by $FV(\phi)$
($FV_{constr}(\phi)$) the set of the free (constraint) variables
of $\phi$.

\begin{definition}
Given an underlying constraint system with set
of constraints  ${\cal C}$,
the  truth of  an atomic formula $X(c)$
is defined with respect to a predicate assignment $v\in  M\rightarrow C$
which assigns to each monadic predicate $X$ a constraint.
We define
\[
\mbox{\rm $v \models  X(c)$ if $v(X)\vdash c$.}\]
\end{definition}

Thus $X(c)$ holds if $c$ is entailed by the constraint
represented by $X$. In other words, a monadic constraint
predicate $X$ denotes a set $\{d\mid\;d\vdash c\}$ for some $c$.
We restrict to constraint predicate assignments which are
monotonic in the following sense: $v(O)\vdash v(I)$. In other
words, the output of a process contains its input.

The temporal operators are interpreted with respect to finite
sequence $\rho=v_1,\ldots,v_n$ of constraint predicate
assignments in the standard manner: $\bigcirc \phi$ holds if
$\phi$ holds in the next time-instant and $\phi \;{\cal
U}\;\psi$ holds if there exists a future moment (possibly the
present) in which $\psi$ holds and until then $\phi$ holds. We
restrict to sequences $\rho=v_1,\ldots,v_n$ which are monotonic
in the following sense: for $1\leq i< n$, we have
\begin{itemize}
\item
$v_{i+1}(X)\vdash v_i(X)$, for every predicate $X$;
\item
$v_{i+1}(I)\vdash v_i(O)$.
\end{itemize}
The latter condition requires that the input of a process contains its output at the previous time-instant. Note
that these conditions corresponds with the monotonicity of reactive sequences as defined above.

In order to define the truth of a temporal  formula we introduce the following notions:  given a finite sequence
$\rho=v_1,\ldots,v_n$ of predicate assignments, we denote by $l(\rho)=n$ the length of $\rho$ and $\rho_i = v_i$,
$1\leq i\leq n$. We also define $\rho< \rho'$ if $\rho$ is a proper suffix of $\rho'$ ($\rho\leq \rho'$ if $\rho<
\rho'$ or $\rho=\rho'$). Given a variable $x$ of the underlying constraint systems and a predicate assignment $v$
we define the predicate assignment $\exists x v$ by $\exists x v(X)=\exists_x d$, where $d=v(X)$. Given a sequence
$\rho=v_1,\ldots,v_n$, we denote by $\exists x\rho$ the sequence $\exists x v_1,\ldots,\exists x v_n$. Moreover,
given a monadic constraint predicate $X$ and a predicate assignment $v$ we denote by $\exists X v$ the {\em
restriction} of $v$ to $M\setminus \{X\}$. Given a sequence $\rho=v_1,\ldots,v_n$, we denote by $\exists X\rho$
the sequence $\exists X v_1,\ldots,\exists X v_n$. Furthermore, by $\gamma$ we denote a constraint assignment
which assigns to each constraint variable $p$ a constraint $\gamma(p)$. Finally, $\gamma\{c/p\}$ denotes the
result of assigning in $\gamma$ the constraint $c$ to the variable $p$.

Moreover, we assume that time does not stop, so actually a finite sequence $v_1\cdots v_n$ represents the
infinite sequence $v_1\cdots v_n,v_n,v_n \cdots$ with the last element repeated infinitely many times. Formally,
this assumption is reflected in the following definition in the interpretation of the $\bigcirc$. By a slight
abuse of notation, given a sequence  $\rho=v_1\cdots v_n$ with $n\geq1$ we define  $\bigcirc\rho$ as follows
\[
\begin{array}{ll}
(n=1) & \bigcirc v_1 = v_1
\\
(n>1) & \bigcirc \rho =v_2\cdots v_n.
\end{array}
\]

The truth of a temporal formula is then defined as follows.

\begin{definition}\label{modelsgamma}
Given a sequence of predicate assignments $\rho = v_1, v_2,\ldots, v_n$, a constraint assignment $\gamma$ and
$\phi$ a temporal formula, we define $\rho \models_\gamma\phi$ by:

by:
\[
\begin{array}{lll}
\rho\models_\gamma p\leq q &\mbox{\rm if}& \gamma(q) \vdash \gamma(p)\\
\rho\models_\gamma X(c)&\mbox{\rm if}& \rho_1 \models X(c)\\
\rho\models_\gamma \neg\phi &\mbox{\rm if}&\rho\not\models_\gamma\phi\\
\rho\models_\gamma \phi_1\wedge \phi_2&\mbox{\rm if}&
\mbox{\rm $\rho   \models_\gamma\phi_1$ and $\rho  \models_\gamma\phi_2$}\\
\rho\models_\gamma\exists x\phi&\mbox{\rm if}& \mbox{\rm $\rho'\models_\gamma \phi$, for some $\rho'$ s.t.
$\exists x \rho=\exists x \rho'$}\\
\rho\models_\gamma\exists X\phi&\mbox{\rm if}& \mbox{\rm $\rho'\models_\gamma \phi$, for some $\rho'$ s.t.
$\exists X \rho=\exists X \rho'$}\\
\rho\models_\gamma\exists p\phi&\mbox{\rm if}& \mbox{\rm $\rho\models_{\gamma'}\phi$, for some $c$ s.t.
$\gamma'=\gamma\{c/p\}$}\\
\rho \models_\gamma\bigcirc \phi&\mbox{\rm if}&
\bigcirc\rho \models_\gamma\phi\\
\rho \models_\gamma \phi\;{\cal U}\; \psi&\mbox{\rm if}& \mbox{\rm for some $\rho'\leq \rho$, $\rho'
\models_\gamma \psi$ and for all $\rho'< \rho'' \leq \rho$, $\rho'' \models_\gamma \phi$}.
\end{array}
\]

Moreover $\rho \models\phi$ iff $\rho \models_\gamma\phi$ for
every constraint assignment $\gamma$.
\end{definition}

\begin{definition}
A formula $\phi$ is valid, notation $\models \phi$, iff
$\rho\models \phi$ for every sequence $\rho$ of predicate
assignments.
\end{definition}

We have the validity of the usual temporal
tautologies.
Monotonicity of the constraint predicates wrt the entailment
relation of the underlying constraint system is
expressed by the formula
\[
\forall p\forall q\forall X
(p\leq q\rightarrow (X(q)\rightarrow X(p))).\]
Monotonicity of the constraint predicates wrt time
implies the validity of the following formula
\[
\forall p\forall X( X(p)\rightarrow \Box X(p)).\]
The relation between the distinguished constraint
predicates $I$ and $O$ is
logically described by the laws
\[
\forall p( I(p)\rightarrow O(p))\;{\rm and}\;
\forall p(O(p)\rightarrow\bigcirc I(p)),
\]
that is, the output of a process contains its input and
is contained in the inputs of the next time-instant.

\subsection{The proof-system}

We introduce now a proof-system for reasoning about the
correctness of {\em tccp} programs. We first define formally the
correctness assertions and their validity.

\begin{definition}\label{corrass}
Correctness assertions are of the form
$ A\;{\it sat}\; \phi$, where $A$ is a {\em tccp} process and $\phi$
is a temporal formula. The validity of an assertion $ A\;{\it sat}\; \phi$,
denoted by  $\models  A\;{\it sat}\; \phi$,
is defined as follows
$$\models A\;{\it sat}\; \phi \hbox{ iff } \rho\models\phi,
\hbox{ for all } \rho \in R'(A), $$ where
\[R'(A)=\{
v_1,\ldots,v_n \mid
\begin{array}[t]{ll}  \la v_1(I),v_1(O)\ra\cdots\la v_n(I),v_n(O)\ra\in
R(A)  &\}. \end{array}
\]
\end{definition}

Roughly, the correctness assertion $ A\;{\it sat}\; \phi$ states
that every sequence $\rho$ of predicate assignments  such that
its `projection' onto the distinguished predicates $I$ and $O$
generates a reactive sequence of $A$, satisfies the temporal
formula $\phi$.

\begin{table*}[th]
\begin{center}
\begin{tabular}{|lll|}  \hline
&\mbox{   }&\mbox{   }
\\
\mbox{\bf T1}& $\hbox{\tell}(c) \;{\it sat}\; O(c)\wedge \forall p
(O(p) \rightarrow \exists q (I(q) \wedge q\sqcup c = p))\wedge
\bigcirc \Box {\it stut} $&
\\
&\mbox{   }&\mbox{   }
\\
\mbox{\bf T2}&
$\frac{\displaystyle
A_i\;{\it sat}\;\phi_i, \forall i\in [1,n]}
{\displaystyle
\sum_{i=1}^n \ask(c_i)\rightarrow A_i\;{\it sat}\;
\bigvee_{i=1}^n\big(
(\bigwedge_{j=1}^n \neg I_j \wedge stut)\; {\cal U}\; (I_i\wedge stut\wedge
\bigcirc \phi_i)\big)
\vee \Box (\bigwedge_{j=1}^n\neg I_j\wedge stut)}$
&
\\
&\mbox{   }&\mbox{   }
\\
\mbox{\bf T3}&
$\frac{\displaystyle
A\;{\it sat}\; \phi\quad B\;{\it sat}\; \psi}
{\displaystyle
\mbox{\bf now $c$ then $A$ else $B$}\;{\it sat}\;
(I(c)\wedge \phi)\vee (\neg I(c)\wedge \psi)}$
&
\\
&\mbox{   }&\mbox{   }
\\
\mbox{\bf T4}& $\frac{\displaystyle A\;{\it sat}\; \phi }
{\displaystyle \exists x A\;{\it sat}\; \exists x(\phi\wedge {\it
loc}(x))\wedge {\it inv}(x)
} $&
\\
&\mbox{   }&\mbox{   }
\\
\mbox{\bf T5}& $\frac{\displaystyle A\;{\it sat}\; \phi\quad
B\;{\it sat}\; \psi} {\displaystyle A\parallel B\;{\it sat}\;
\exists X,Y(\phi[X/O]\wedge \psi[Y/O]\wedge {\it par}(X,Y))} \quad
X,Y \not \in FV(\phi)\cup FV(\psi), \ X \neq Y $ &
\\
&\mbox{   }&\mbox{   }
\\
\mbox{\bf T6}& $\frac{\displaystyle p(x)\;{\it sat}\;
\phi\vdash_p A\;{\it sat}\;\phi} {\displaystyle p(x)\;{\it
sat}\;\phi} \quad \mbox{\rm $p(x)$ declared as $A$}$&
\\
&\mbox{   }&\mbox{   }
\\
\mbox{\bf T7}& $\frac{\displaystyle A\;{\it sat}\; \phi\quad
\models\phi\rightarrow\psi} {\displaystyle A\;{\it sat}\;\psi}$ &
\\
&\mbox{   }&\mbox{   }
\\
\hline
\end{tabular}
\caption{The system TL for ${\it tccp}$.}\label{table:TL}
\end{center}
\end{table*}

Table \ref{table:TL}
presents the proof-system.

Axiom {\bf T1} states that the execution of $\hbox{\tell}(c)$
consists of the output of $c$ (as described by $O(c)$) together
with any possible input (as described by $I(q)$). Moreover, at
every time-instant in the future no further output is generated,
which is expressed by the formula
\[\forall p
(O(p)\leftrightarrow I(p)),
\]
which we abbreviate by {\it stut}
(since it represents  stuttering steps).

In rule {\bf T2}  $I_i$ stands for $I(c_i)$.
Given that $A_i$ satisfies $\phi_i$, rule
{\bf T2} allows the derivation of the specification
for $\Sigma_{i=1}^n \ask(c_i)\rightarrow A_i$,
which expresses that either
eventually $c_i$ is an input and, consequently,
$\phi_i$ holds in the {\em next} time-instant
(since the evaluation of the ask takes one time-unit), or
none of the guards is ever satisfied.

Rule {\bf T3} simply states that if $A$ satisfies $\phi$ and $B$
satisfies $\psi$ then every computation of $\mbox{\bf now $c$
then $A$ else $B$}$ satisfies either $\phi$ or $\psi$, depending
on the fact that $c$ is an input or not.

Hiding of a local variable $x$ is axiomatized in rule {\bf T4} by first
existentially quantifying $x$ in
$\phi\wedge {\it loc}(x)$, where ${\it loc}(x)$
denotes  the following formula which expresses
that $x$ is local, i.e.,
the inputs of the environment do not contain
new information on $x$:
\[
\forall p (\exists_x p \neq p \rightarrow (\neg I(p) \wedge
            \Box (\bigcirc I(p) \rightarrow \exists r
            (O(r) \wedge \exists_x p \sqcup r = p ) ) ) ).
            \]
This formula literally states that the initial input does not
contain information on $x$ and that everywhere in the computation
if in the next state an input contains information on $x$ then
this information is already contained by the previous output.
Finally, the following formula ${\it inv}(x)$
\[
 \forall p\Box  (\exists_x p \neq p \rightarrow
              (O(p) \rightarrow \exists r
                   (I(r) \wedge \exists_x p \sqcup r = p ) ) )
            \]
states that the process does not provide new information
on the {\it global} variable $x$.

Rule {\bf T5} gives a compositional axiomatization of parallel
composition. The `fresh' constraint predicates $X$ and $Y$ are
used to represent the outputs of $A$ and $B$, respectively
($\phi[X/O]$ and $\psi[Y/O]$ denote the result of replacing $O$
by $X$ and $Y$). Additionally, the formula
\[
\forall p\Box
(O(p)\leftrightarrow (\exists q_1,q_2 (X(q_1)\wedge Y(q_2) \wedge q_1\sqcup q_2
 = p))),\]
denoted by ${\it par}(X,Y)$, expresses that
every output of $A\parallel B$ can be decomposed
into outputs of $A$ and $B$.

Rule
{\bf T6}, where $\vdash_p$ denotes derivability within
the proof system, describes recursion in the usual manner
by using a meta-rule (Scott-induction, see also
\cite{BGMP94}): we can conclude that the agent
$p(x)$ satisfies a property $\phi$ whenever the body of
$p(x)$ satisfies the same property assuming the conclusion of the rule.
In this rule $x$ is assumed to be both the
formal and the actual parameter. We do not need more rules since,
as previously mentioned, we can assumed that the set {\it D} of
procedure declarations is closed wrt parameter names.

Note also that, for the sake of simplicity, we do not mention
explicitly the declarations in the proof system. In fact, the more
precise formulation of this rule, that will be needed in the
proofs, would be the following:
$$
\frac{\displaystyle D\setminus \{p\}. p(x)\;{\it sat}\;
\phi\vdash_p D\setminus \{p\}. A\;{\it sat}\;\phi}
{\displaystyle D.p(x)\;{\it sat}\;\phi}.
$$

Rule {\bf T7} allows to weaken the
specification.

As an example of  a sketch of a derivation consider the agent
$\exists x A$ where
$$ \begin{array}{ll}
A::  & \ask(x=a)\rightarrow \hbox{\tell}(true)
\\
      & +
\\
      & \ask(true)\rightarrow \hbox{\tell}(y=b).
\end{array}
$$
(constraints are equations on the Herbrand universe).
By {\bf T1} and {\bf T7} we derive
\[
\hbox{\tell}(y=b)\;{\it sat}\; O(y=b)\quad{\mbox\rm and}\quad
\hbox{\tell}(true)\;{\it sat}\; O(true).\]

By {\bf T2} and {\bf T7} we subsequently derive
$$A\;{\it sat}\; I(x=a)\vee \bigcirc O(y=b)$$
(note that $\neg I(\true)$ is logically equivalent to ${\it false}$ and ${\it false}\;{\cal U}\; \phi$ is
equivalent to $\phi$). Using rule {\bf T4}, we   derive the correctness assertion
$$\exists x A\;{\it sat}\; \exists x( (I(x=a)\vee \bigcirc O(y=b))\wedge
{\it loc}(x)).$$ It is easy to see that $I(x=a)\wedge {\it
loc}(x)$ implies {\it false}. So we have that $\exists x(
(I(x=a)\vee \bigcirc O(y=b))\wedge {\it loc}(x))$ implies $\exists
x ({\it loc}(x)\wedge \bigcirc O(y=b))$. Clearly this latter
formula implies $\bigcirc O(y=b)$. Summarizing the above, we
obtain a derivation of the correctness assertion $\exists xA\;{\it
sat}\; \bigcirc O(y=b)$ which states that in every reactive
sequence of $\exists xA$ the constraint $y=b$ is produced in the
next (wrt the start of the sequence) time instant.

\section{Soundness and completeness}\label{sec:densem}

We investigate now soundness and completeness of the above
calculus. Here and in the following, in order to clarify some
technical details, we consider processes of the form $D.A$ rather
than agents of the form $A$ with a separate set of declarations
$D$. All the previous definitions can be extended to processes in
the obvious way. We also denote by $\vdash_p D.A\;{\it sat}\;\phi$
the derivability of the correctness assertion $D.A\;{\it
sat}\;\phi$ in the proof system introduced in the previous section
(assuming as additional axioms in rule {\bf T7} all valid temporal
formulas).

Soundness means that every provable correctness assertion is
valid: whenever $\vdash_p D.A \;sat\; \phi$, i.e. $D.A \;sat\;
\phi$ is derivable, then $\models D.A \;sat\; \phi$. Completeness
on the other hand consists in the derivability of every valid
correctness assertion: whenever $\models D.A \;sat\; \phi$ then
$\vdash_p D.A\;sat\; \phi$ (in $T$).

At the heart of the soundness and completeness results that we are going to prove  lies the compositionality of
the semantics $R'$, which follows from the compositionality of the underlying semantics $R$. In order to prove
such a compositionality, we first introduce a denotational semantics $\os D.A \cs(e)$ where, for technical
reasons, we represent explicitly the environment $e$ which associate a denotation to each procedure identifier.
More precisely, assuming that {\it Pvar} denotes the set of procedure identifier, ${\it Env}={\it Pvar}\rightarrow
\wp({\cal S})$, with typical element $e$, is the set of {\it environments}.

Given a process $D.A$, the denotational semantics $\os D.A \cs
:{\it Env} \rightarrow \wp({\cal S})$ is defined by the equations
in Table~\ref{densem}, where $\mu$ denotes the least fixpoint wrt
subset inclusion of elements of $\wp({\cal S})$. The semantic
operators appearing in Table~\ref{densem} are formally defined as
follows. Intuitively they reflect, in terms of reactive sequences,
the operational behaviour of their syntactic counterparts.

\begin{definition}\cite{BGM00}\label{defsemoperators}
Let $S,S_i$ be sets of reactive sequences and
 $c,c_i$ be constraints.
Then we define the operators $\tilde{\sum}$, $\tilde{\parallel}$,
$\tilde{now}$ and $\tilde{\exists} x$ as follows:

\noindent
{\bf Guarded Choice}
\[
{\bf \tilde{\sum} } _{i=1}^n c_i \rarrow S_i =
\begin{array}[t]{l}
\{ s\cdot s'   \in {\cal S} \mid
\begin{array}[t]{l}
s = \la d_1 ,d_1\ra \cdots \la d_m,d_m\ra, d_j \not\vdash c_i
\mbox{ for each } j\in [1,m$-$1], i\in [1,n],
\\
d_m \vdash c_h \hbox { and } s'\in S_h  \hbox{ for an
} h\in[1,n] \ \}
\end{array}
\\
\cup
\\
\{ s \in {\cal S}  \mid
\begin{array}[t]{l}
s = \la d_1,d_1\ra \cdots \la d_m,d_m\ra, d_j \not\vdash c_i
\mbox{ for each } j\in [1,m], \  i\in [1,n]  \},
\end{array}
\end{array}
\]

\noindent
{\bf Parallel Composition}
Let $\tilde{\parallel} \in{\cal S}\times
{\cal S}\rightarrow {\cal S}$ be the
(commutative and associative) partial operator defined as follows:
\[
\begin{array}{l}
\la c_1,d_1\ra \cdots\la c_n,d_n\ra \la d,d\ra
\tilde{\parallel}
\la c_1,e_1\ra\cdots\la c_n,e_n\ra  \la d,d\ra=
\la c_1, d_1\sqcup e_1\ra\cdots
\la c_n, d_n\sqcup e_n \ra\la d,d\ra.
\end{array}
\]
We define $S_1\tilde{\parallel} S_2$ as the point-wise extension
of the above operator to sets.

\noindent
{\bf The Now-Operator}
\[
\tilde{now}(c,  S_{1} , S_{2})=
        \{ s  \in {\cal S} \mid
        s = \la c' ,d \ra \cdot s' \hbox{ and either } \\
        c' \vdash c \mbox{ and } s \in S_{1} \hbox{ or }
        c' \not \vdash c \mbox{ and } s \in S_{2}\ \}.
\]

\noindent
{\bf The Hiding Operator}
We first need the following notions similar to those used in
\cite{BKPR92}:

Given a sequence $s = \la c_1,d_1\ra \cdots \la c_n,c_n\ra$, we
denote by $\exists x s$ the sequence $\la \exists x c_1,\exists x
d_1\ra \cdots \la \exists x c_n,\exists x c_n\ra$.

A sequence $s' = \la c_1,d_1\ra \cdots \la c_n,c_n\ra$ is
$x$-connected if
\begin{itemize}
\item $\exists_x c_1 = c_1$ (that is, the input constraint of $s'$ does
not
contain information on $x$) and
\item $\exists_x c_i \sqcup d_{i-1} = c_i \hbox{ for each } i\in[2,n]$
(that is, each assumption $c_i$ does not contain any information on $x$
which has not been produced previously in the sequence by some $d_j$).
\end{itemize}

A sequence $s$ is $x$-invariant  if
\begin{itemize}
\item for all computation steps $\la c,d\ra$ of $s$, $d=\exists_x d\sqcup
c$ holds.
\end{itemize}
The semantic hiding operator then can be defined as follows:
\[
{\bf \tilde{ \exists} } x S =  \{ s \in {\cal S} \mid \mbox{\rm there exists $ s' \in S$ such that $ \exists_x s
= \exists_x s'$, $s'$ is $x$-connected and $s$ is $x$-invariant} \}.
\]
\end{definition}

A few explanations are in order here.
Concerning the semantic choice operator, a
sequence in ${\bf \tilde{\sum} } _{i=1}^n c_i \rarrow S_i$
consists of an initial period
of waiting for (a constraint stronger than) one of the constraints $c_i$.
During this waiting period only the environment is active by producing the
constraints $d_i$ while the process itself generates the stuttering steps
$\langle d_i,d_i\rangle$. Here we can add several pairs since
the external environment can take several time-units to produce the
required constraint.  When the contribution of the environment is strong
enough to entail a $c_h$ the resulting sequence is  obtained by adding
$s'\in S_h$ to
the initial waiting period.

In the semantic parallel operator defined on sequences
we require that the two arguments of the operator agree
at each point of time with respect to the contribution of the environment
(the $c_i$'s) and that they have the same length (in all other cases the
parallel composition is assumed being undefined).

In the definition of $\tilde{\exists}$ we say
that a sequence is $x$-connected if  no information on $x$ is
present in the input constraints which has not been already
accumulated by the computation of the agent  itself.
A sequence is $x$-invariant if its
computation steps do not provide more information on $x$.

If $D.A$ is a closed process, that is if all the procedure names occurring in $A$ are defined in $D$, then  $\os
D.A\cs(e)$ does not depend on $e$ and will be indicated as $\os D.A\cs$. Environments in general allows us to
defined the semantics also of processes which are not closed, and this will be used in the soundness proof.

The following result shows the correspondence between the two
semantics we have introduced and therefore the compositionality
of $R(A)$.

\begin{theorem}\label{reqden} \cite{BGM00}
If $D.A$ is closed then $R(A) = \os D.A \cs$ holds.
\end{theorem}

\begin{table*}[th]
\begin{center}
\begin{tabular}{| l  l  l|}\hline

\mbox{ }&\mbox{ }&\\

\mbox{{\bf E1}} & $\os {\it D.{\bf stop}} \cs(e) = \{ \la c_1,
c_1\ra \la c_2, c_2\ra  \cdots \la c_n, c_n\ra \in {\cal S}
\mid n\geq 1\}$&\\

\mbox{ }&\mbox{ }&\\

\mbox{{\bf E2}} & $\os {\it D.\hbox{\tell}(c)} \cs(e) = \{ \la d,
d\sqcup c\ra \cdot s \in {\cal S}
\mid s\in \os D.{\bf stop} \cs(e) \}$&\\

\mbox{ }&\mbox{ }&\\

\mbox{{\bf E3}} & $\os {\it D.\sum_{i=1}^n }{\it \ask(c_i)\rarrow
A_i} \cs(e) =
\tilde{\sum}_{i=1}^n c_i\rightarrow \os {\it D.A_i} \cs(e)$&\\

\mbox{ }&\mbox{ }&\\

\mbox{{\bf E4}} & $\os {\it D.{\bf now}\ c\  {\bf then} \ A \ {\bf
else} \ B}\cs(e)=
\tilde{{\it now}}(c,\os{\it D.A}\cs(e),\os {\it D.B}\cs(e))$&\\

\mbox{ }&\mbox{ }&\\

\mbox{{\bf E5}} & $\os {\it D.A\parallel B}\cs(e) = \os {\it
D.A}\cs(e) \ {\it \tilde{ \parallel } }
\ \os {\it D.B}\cs(e)$&\\

\mbox{ }&\mbox{ }&\\

\mbox{{\bf E6}} &
$\os {\it D.\exists x A}\cs(e) = \tilde{\exists}x \os{\it D.A} \cs(e)$&\\

\mbox{ }&\mbox{ }&\\

\mbox{{\bf E7}} & $\os {\it D.p(x)} \cs(e) =
\mu \Psi$&\\
&
where
$\Psi(f)=\os D\setminus\{p\}.A\cs (e\{f/p\})$,
$p(x)::A\in D$&\\
\mbox{ }&\mbox{ }&\\
\hline
\end{tabular}
\end{center}
\caption{The semantics  $\os D.A \cs$(e).}\label{densem}
\end{table*}

In order to prove the soundness of the calculus we have to
interpret also correctness assertions about arbitrary processes
(that is, including processes which do contain undefined procedure
variables).

\begin{definition}\label{modelse}
Given a underlying constraint system ${\cal C}$
and an environment $e$, we define
$\models_e D.A \;sat\; \phi$ iff
$$\rho\models\phi,
\hbox{ for all } \rho \in \os  D.A \cs' (e),
$$ where
\[\os D.A \cs' (e) =\{
v_1,\ldots,v_n \mid \begin{array}[t]{ll}\la v_1(I),v_1(O)\ra\cdots\la v_n(I),v_n(O)\ra\in \os D.A \cs (e) & \}.
\end{array}
\]
\end{definition}

Note that, for closed processes, $\models_e$ coincides with
$\models$ as previously defined. We first need the following
Lemma.

\begin{lemma}\label{daungamma}
Let $\phi$ be a temporal formula, $\rho$ and $\rho'$ be sequences
of predicate assignment and let $V$ be a variable such that either
$V \in M$ or $V$ is a variable $x$ of the underlying constraint
system. The following holds:
\begin{enumerate}
  \item
  Assume that $\rho \models \phi$. Then $\rho' \models \exists V \phi$
  for each $\rho'$ such that $\exists V \rho=\exists V \rho'$.
  \item Assume that $\rho \models \exists V \phi$ and
  $FV_{constr}(\phi)= \emptyset$. Then there exists $\rho'$
  such that $\rho' \models \phi$ and
  $\exists V \rho' = \exists V \rho$.
\end{enumerate}
\end{lemma}
{\bf Proof}
\begin{enumerate}
  \item Assume that $\rho \models \phi$. By Definition~\ref{modelsgamma},
  $\rho \models_\gamma \phi$ for each $\gamma$ and then for each
  $\rho'$ such that $\exists V \rho=\exists V \rho'$,
  we have that $\rho'\models_\gamma \exists V \phi$ for each
  $\gamma$. Therefore, by Definition~\ref{modelsgamma},
  $\rho' \models  \exists V \phi$.
  \item Assume that $\rho \models \exists V \phi$ and
  $FV_{constr}(\phi)= \emptyset$. By
  Definition~\ref{modelsgamma},
  $\rho \models_\gamma \exists V \phi$ for each $\gamma$.
  Then by Definition~\ref{modelsgamma}, for each $\gamma$
  there exists $\rho'$ such that $\exists V \rho=\exists V \rho'$
  and $\rho' \models_\gamma \phi$.
  Since by hypothesis $FV_{constr}(\phi)= \emptyset$, whenever
  $\rho' \models_\gamma \phi$ we also have that
  $\rho' \models \phi$ and then the thesis holds.
\end{enumerate}

The following Theorem is the core of the soundness result.

\begin{theorem}\label{exsoundness}
Let us denote $D_i.A_i$ by $P_i$ for $1\leq i\leq n$ and $D.A$ by
$P$. If
$$P_1\;sat\;\phi_1,\ldots,P_n\;sat\;\phi_n \vdash_p
P\;sat\;\phi \hbox{ and } \models_e P_i\;sat\;\phi_i,$$ for $i=1,\ldots,n$, then
$$\models_e P\;sat\;\phi.$$
\end{theorem}
{\bf Proof} The proof is by induction on the length $l$ of the
derivation.
\begin{description}
  \item[($l=1$)] In this case $A=\hbox{\tell}(c)$ and $\vdash_p D.\hbox{\tell}(c)
   \;{\it sat}\; O(c)\wedge \forall p (O(p) \rightarrow
   \exists q (I(q) \wedge q\sqcup c = p))\wedge
   \bigcirc \Box {\it stut} $.
   By Definition~\ref{modelse}, we have to prove that for any $e$,
   $\rho\models O(c)\wedge \forall p (O(p) \rightarrow
   \exists q (I(q) \wedge q\sqcup c = p))\wedge
   \bigcirc \Box {\it stut} $, for all
   $\rho \in \os  D.\hbox{\tell}(c)\cs' (e).$
   By equation {\bf E2} of Table~\ref{densem} and Definition~\ref{modelse},
   \[ \os  D.\hbox{\tell}(c)\cs' (e)=\{\rho \mid \rho_1(O)=\rho_1(I)\sqcup
   c \mbox{ and } \rho_i(O)=\rho_i(I) \mbox{ for } i \in [2,l(\rho)]
   \}.\]
   The remaining of the proof for this case is straightforward.
  \item[($l>1$)] We distinguish various cases according to the last
  rule applied in the derivation.
\begin{description}
  \item[Rule {\bf T2}.]
  In this case $\vdash_p D.\sum_{i=1}^n \ask(c_i)\rightarrow A_i
   \;{\it sat}\; \chi$, where $\chi$ is the formula
  \[\bigvee_{i=1}^n\big( (\bigwedge_{j=1}^n \neg I(c_j) \wedge stut)\;
  {\cal U}\; (I(c_i)\wedge stut\wedge \bigcirc \phi_i)\big) \vee \Box
  (\bigwedge_{j=1}^n\neg I(c_j)\wedge stut).\]
  Since for each $i\in [1,n]$ the proof  $D.A_i\;{\it sat}\;\phi_i$ is
  shorter than the current one, from the inductive hypothesis follows
  that, for every environment $e$ and for each $i\in [1,n]$,
  $\models_e D.A_i\;{\it sat}\;\phi_i$, that is
  $\rho\models \phi_i$, for all $\rho \in \os  D.A_i \cs' (e).$

  Let us take a particular $e$. By Definition~\ref{modelse},
  we have to prove that $\rho\models\chi$, for all
  $\rho \in \os  D.\sum_{i=1}^n \ask(c_i)\rightarrow A_i \cs' (e).$

  By equation {\bf E3} of Table~\ref{densem} and
  Definitions~\ref{defsemoperators} and \ref{modelse},
  $\os  D.\sum_{i=1}^n \ask(c_i)\rightarrow A_i \cs'(e)=D_1 \cup D_2$, where
  \[
  \begin{array}{lll}
  D_1 = & \{\rho \cdot \rho' \mid & \rho_j(O)=\rho_j(I)
  \mbox{ and } \rho_j(I)\not\vdash c_i
  \mbox{ for each } j\in [1,l(\rho)-1], i\in [1,n] \\
  &  & \rho_{l(\rho)}(I)\vdash c_h, \
  \rho_{l(\rho)}(O)=\rho_{l(\rho)}(I)
  \hbox { and } \rho'\in \os  D.A_h \cs' (e)
  \hbox{ for an } h\in[1,n]\}  \mbox{ and }
  \\
  D_2 = & \{\rho \mid & \rho_j(O)=\rho_j(I) \mbox{ and }
  \rho_j(I)\not\vdash c_i
  \mbox{ for each } j\in [1,l(\rho)], i\in [1,n]\}
  \end{array}
  \]
  Now, it is straightforward to prove that if $\rho \in D_2$ then
  $\rho \models\Box (\bigwedge_{j=1}^n\neg I(c_j)\wedge
  stut)$.
  Moreover, since by inductive hypothesis for each $i\in [1,n]$,
  $\models_e D.A_i\;{\it sat}\;\phi_i$, we
  have that if $\rho \in D_1$ then
  $\rho \models \bigvee_{i=1}^n\big( (\bigwedge_{j=1}^n \neg I(c_j) \wedge stut)\;
  {\cal U}\; (I(c_i)\wedge stut\wedge \bigcirc \phi_i)\big)$ and
  then the thesis.
  \item[Rule {\bf T3}]
  In this case $\vdash_p \mbox{\bf now $c$ then $A$ else $B$}
   \;{\it sat}\; (I(c)\wedge \phi)\vee (\neg I(c)\wedge \psi)$.
  Since the proofs  $A\;{\it sat}\;\phi$ and $B\;{\it sat}\;\psi$ are
  shorter than the current one, the induction hypothesis says
  that for every environment $e$, we have that
  $\models_e D.A\;{\it sat}\;\phi$ and $\models_e D.B\;{\it sat}\;\psi$ i.e.
  $\rho\models\phi$ and $\rho'\models\psi$ for all
  $\rho \in \os  D.A \cs' (e)$ and $\rho' \in \os  D.B\cs' (e).$

  Let us take a particular $e$.
  By Definition~\ref{modelse}, we have to prove that
  $\rho\models (I(c)\wedge \phi)\vee (\neg I(c)\wedge \psi)$ for all
  $\rho \in \os D. \mbox{\bf now $c$ then $A$ else $B$} \cs'(e)$.
  Now the proof is immediate, by observing that by equation {\bf E4}
  of Table~\ref{densem} and
  Definitions~\ref{defsemoperators} and \ref{modelse},
  $\os D. \mbox{\bf now $c$ then $A$ else $B$} \cs'(e)=
  D_1 \cup D_2$, where
  \[
  \begin{array}{lll}
   D_1 = & \{\rho \mid \rho_1(I)\vdash c  \mbox{ and } \rho\in \os  D.A\cs'(e)\}
    \mbox{ and }
   \\
   D_2 = & \{\rho' \mid \rho'_1 (I)\not\vdash c \hbox{ and } \rho'\in \os  D.B \cs'
   (e)\}.
  \end{array}
  \]
  \item[Rule {\bf T4}]
  In this case $\vdash_p \exists x A\;{\it sat}\;
  \exists x(\phi\wedge {\it loc}(x))\wedge {\it inv}(x)$.
  Since the proof $A\;{\it sat}\;\phi$ is
  shorter than the current one, the induction hypothesis says
  that for every environment $e$,
  $\models_e D.A\;{\it sat}\;\phi$ i.e.
  $\rho'\models\phi$ for all $\rho' \in \os  D.A \cs' (e)$.
  Let us consider a particular $e$.
  By Definition~\ref{modelse}, we have to prove that $\rho\models
  \exists x(\phi\wedge {\it loc}(x))\wedge {\it inv}(x)$, for all
  $\rho \in \os D. \exists x A\cs' (e).$
  By equation {\bf E6} of Table~\ref{densem} and
  Definitions~\ref{defsemoperators} and \ref{modelse},
  $\rho \in \os D. \exists x A \cs'(e)$ if and only if there exists
  $\rho'\in \os D.A \cs'(e)$ such that $l(\rho)=l(\rho')$ and the
  following conditions hold
\begin{enumerate}
  \item for each $i\in [1,l(\rho)]$,
  $\exists x \rho_i(I)=\exists x \rho'_i(I)$ and $\exists x \rho_i(O)=\exists x
  \rho'_i(O)$,
  \item $\exists x \rho'_1(I)=\rho'_1(I)$ and for each $i\in [2,l(\rho)]$,
  $\exists x \rho_i'(I)\sqcup \rho'_{i-1}(O)=\rho'_i(I)$,
  \item for each $i\in [1,l(\rho)]$,
  $\rho_i(O) = \exists x \rho_i(O)\sqcup \rho_{i}(I)$.
\end{enumerate}
Since for any sequence $\rho$ of predicate assignments and for
any tccp process $D.A$, $\rho \in \os D.A\cs' (e)$  if and only
if its `projection' onto the distinguished predicates $I$ and $O$
generates a reactive sequence of $D.A$, by 1. we can assume
without loss of generality that  $\exists x \rho=\exists x
\rho'$. From 2., the definition of $loc(x)$ and the inductive
hypothesis follows that $\rho'\models \phi\wedge {\it loc}(x)$.
Therefore, by the previous equality and case 1 of
Lemma~\ref{daungamma} imply that $\rho\models \exists
x(\phi\wedge {\it loc}(x))$ holds. Moreover by 3. and by
definition of $inv(x)$ we obtain that $\rho\models inv(x)$ thus
proving the thesis for this case.
  \item[Rule {\bf T5}]
  In this case $\vdash_p A\parallel B\;{\it sat}\; \chi$, where
  $\chi$ is the formula
  \[\exists X,Y(\phi[X/O]\wedge \psi[Y/O]\wedge {\it par}(X,Y)),\]
  $X,Y \in M$, $X \neq Y$ and $X,Y \not \in FV(\phi)\cup FV(\psi)$.
  Since the proofs  $A\;{\it sat}\;\phi$ and $B\;{\it sat}\;\psi$ are
  shorter than the current one, the induction hypothesis says
  that for every environment $e$, we have that
  $\models_e D.A\;{\it sat}\;\phi$ and $\models_e D.B\;{\it sat}\;\psi$ i.e.
  $\rho'\models \phi$ and $\rho''\models \psi$ for all
  $\rho' \in \os  D.A \cs' (e)$ and $\rho'' \in \os  D.B\cs' (e).$

  Let us take a particular $e$.
  By Definition~\ref{modelse}, we have to prove that $\rho\models \chi$,
  for all $\rho \in \os D. A\parallel B\cs' (e).$
  Assume that $\rho \in \os D. A\parallel B \cs'(e)$.
  By equation {\bf E5} of Table~\ref{densem} and
  Definitions~\ref{defsemoperators} and \ref{modelse},
  there exist $\rho'\in \os D.A \cs'(e)$ and
  $\rho'' \in \os D.B \cs'(e)$ such that $l(\rho)=l(\rho')=l(\rho'')$ and
  for each $i\in [1,l(\rho)]$, $\rho_i(I)=\rho'_i(I)=\rho''_i(I)$ and
  $\rho_i(O)=\rho'_i(O) \sqcup \rho'_i(O).$
  Since for any sequence $\rho$ of predicate assignments and for any
  tccp process $D.A$, $\rho \in \os D.A\cs' (e)$  if and only if its
  `projection' on the distinguished predicates $I$ and $O$
  generates a reactive sequence of $D.A$,  by previous observation we can
  assume without loss of generality that
 $\rho_i(Z)=\rho'_i(Z)=\rho''_i(Z)$ for each $i\in [1,n]$ and for each $Z \in M$
 such that $Z\neq O$.
  Now we can construct a new sequence $\bar \rho$ of predicate
  assignments, of the same length of $\rho$, such that
  $\bar\rho_i(X)=\rho'_i(O)$
  $\bar\rho_i(Y)=\rho''_i(O)$
  $\bar\rho_i(Z)=\rho_i(Z)$ for each $i\in [1,l(\rho)]$
   and for each $Z \in M$ such that $Z \neq X,Y$.
  Since $X$ and $Y$ are not free in $\phi$ and $\psi$ and by inductive hypothesis
  $\rho'\models\phi$ and $\rho''\models\psi$ holds,
  by construction we obtain that $\bar\rho\models\phi[X/O]$ and
  $\bar\rho\models\psi[Y/O]$.
  Moreover, by construction $\bar\rho\models{\it par}(X,Y)$ holds.
  Therefore $\bar\rho\models\phi[X/O]\wedge \psi[Y/O]\wedge {\it
  par}(X,Y)$ and the thesis follows from case 1 of Lemma~\ref{daungamma}
  by observing that $\exists X,Y \rho= \exists X,Y \bar \rho$.
  \item[Rule {\bf T6}]
  In this case $\vdash_p D.p(x)\;{\it sat}\;\phi$.
  Since the proof
  $D\setminus\{p\}.p(x)\;sat\; \phi \vdash_p D\setminus\{p\}.A\;sat\; \phi$
  is shorter than the current one, the induction hypothesis says that
  for every environment $e$ such that
  $\models_e D\setminus\{p\}.p(x)\;sat\; \phi$ we also have that
  $\models_e D\setminus\{p\}.A\;sat\; \phi$.
  Let us take a particular $e$.
  We have to show that $\models_e D.p(x)\;\sat\;\phi$ holds or,
  in other words, that if $\rho \in \os D.p(x) \cs(e)$ then
  we have that $\rho \models \phi$.

  Now $\os D.p(x)\cs(e)=\mu\Psi$, where $\mu\Psi=\bigcup_i f_i$,
  with $f_0= \emptyset$ and
  $f_{i+1}=\os D\setminus\{p\}.A\cs(e\{f_i/p\})$.
  Thus it suffices to prove by induction that for all $n$,
  if $ \rho \in f_n$ then, $\rho\models\phi$.
  The base case is obvious. Suppose that the thesis holds for $f_n$,
  so for $e'=e\{f_n/p\}$ we have that
  $\models_{e'} D\setminus\{p\}.p(x)\;\sat\;\phi$, holds.
  Thus we infer that $\models_{e'} D\setminus\{p\}.A\;\sat\;\phi$.
  Since by definition
  $f_{n+1}=\os D\setminus\{p\}.A\cs(e\{f_n/p\})$, we have that if
  $\rho \in f_{n+1}$ then
  $\rho\models \phi$ which completes the proof.

  \item[Rule {\bf T7}] The proof is immediate.

\end{description}
\end{description}

\bigskip

Since $R(A) = \os D.A\cs (e)$ holds for any closed process $D.A$
with $e$ arbitrary, whenever  $\models_e D.A \;sat\; \phi$ we also
have $\models D.A \;sat\; \phi$. Hence from the above result, with
$n=0$, we can derive immediately the soundness of the system.

\begin{corollary}[Soundness]
The proof system consisting of the rules {\bf C0-C7} is sound,
that is, given a closed process $D.A$, if $\vdash_p A \ sat \
\phi$ then $\models D.A\ sat \ \phi$ holds, for every correctness
assertion $D.A \ sat \ \phi$.
\end{corollary}

Following the standard notion of completeness for Hoare-style
proof systems as introduced by \cite{Co78} we consider here a
notion of relative completeness. We assume the existence of a
property which describes exactly the denotation of a process, that
is, we assume that for any process $D.A$ there exists a formula,
that for the sake of simplicity we denote by $\psi(A)$, such that
$\rho \in R'(A)$ iff $\rho \models \psi(A)$ holds\footnote{In
order to describe recursion, the syntax of the temporal formulas
has to be extended with a fixpoint operator of the form $\mu
p(x).\phi$, where $p(x)$ is supposed to occur positively in $\phi$
and the variable $x$ denotes the formal parameter associated with
the procedure $p$ (see \cite{BGMP94}).

The meaning of $\mu p(x).\phi$ is given by a least
fixpoint-construction which is defined in terms of the lattice of
sets of sequences of predicate assignements ordered by
set-inclusion.}. This is analogous to assume the expressibility of
the strongest postcondition of a process $P$, as with standard
Hoare-like proof systems. Furthermore, we assume as additional
axioms all the valid temporal formulas, (for use in the
consequence rule). Also this assumption, in general, is needed to
obtain completeness of Hoare logics.

Analogously to the previous case, the completeness of the system
is a corollary of the following Theorem.

\begin{theorem}
Let $D =\{ p_1(x_1)::A_1,\ldots,p_n(x_n)::A_n \}$ be a set of
declarations and $A$ an agent which involves only calls to
procedures declared in $D$. Then we have
\[
\Phi_1,\ldots,\Phi_n \vdash_p A\;sat\; \psi(A).
\]
where, for $i =1,\ldots,n$, $\Phi_i = p_i(x_i)\;sat\ \psi(p_i(x_i)); $.
\end{theorem}
{\bf Proof} First observe that, for $i =1,\ldots,n$, we can
assume $FV_{constr}(\psi(p_i(x_i)))=\emptyset$. In fact, from the
definition of $\models$ it follows that $\rho \in R'(p_i(x_i))$
if and only if $\rho \models_\gamma \psi(p_i(x_i))$, for each
constraint assignment $\gamma$, and this holds if and only if
$\rho \models \forall_{constr} \psi(p_i(x_i))$, where
$\forall_{constr} \psi(p_i(x_i))$ is the universal closure over
constraint variables of the formula $\psi(p_i(x_i))$. Therefore
we can assume that all the constraint variables in
$\psi(p_i(x_i))$ are universally quantified, thus there are no
free constraint variables.

We prove now, by induction on the complexity of $A$, that
$\Phi_1,\ldots,\Phi_n \vdash_p A\;sat\; \psi(A)$ and
$FV_{constr}(\psi(A))=\emptyset$.

\begin{description}
  \item[{\bf ($\hbox{\tell}(c)$)}] In this case, since
  $FV_{constr}(O(c)\wedge \forall p (O(p)
  \rightarrow \exists q (I(q) \wedge q\sqcup c = p))\wedge \bigcirc \Box {\it
  stut})=\emptyset$,
  we have only to prove that
  \[\psi(\hbox{\tell}(c))= O(c)\wedge \forall p (O(p)
  \rightarrow \exists q (I(q) \wedge q\sqcup c = p))\wedge \bigcirc \Box {\it stut}.
  \]
  The proof is straightforward, since Definition~\ref{corrass},
  Theorem~\ref{reqden} and
  equation~{\bf E2} of Table~\ref{densem} imply that the
  following equalities hold
  \[R'(\hbox{\tell}(c))=\{\rho \mid \rho_1(O)=\rho_1(I)\sqcup c \mbox{ and }
  \rho_i(O)=\rho_i(I) \mbox{ for } i \in [2,l(\rho)] \}.\]

\item[{\bf ($\Sigma_{i=1}^n \ask(c_i)\rightarrow A_i$)}] By inductive hypothesis we obtain that
  \[\Phi_1,\ldots,\Phi_n \vdash_p A_i\;sat\;\psi(A_i)
  \mbox{ and } FV_{constr}(\psi(A_i))=\emptyset, \]
  for $i=1,2,\ldots,n$. Then by rule {\bf T2} we obtain also that
  \[\Phi_1,\ldots,\Phi_n \vdash_p A\;sat\; \beta\]
  where
  \[\beta = \bigvee_{i=1}^n\big(
(\bigwedge_{j=1}^n \neg I_j \wedge stut)\; {\cal U}\; (I_i\wedge stut\wedge \bigcirc \psi (A_i))\big) \vee \Box
(\bigwedge_{j=1}^n\neg I_j\wedge stut),\].

  The inductive hypothesis implies that that
 $$ FV_{constr}(\beta)=\emptyset.
 $$

  Then, in order to prove the thesis we have to show that
  \[\beta
  = \psi(\Sigma_{i=1}^n \ask(c_i)\rightarrow A_i)\] holds,
  that is, we have to prove that $\rho \in R'(\Sigma_{i=1}^n \ask(c_i)\rightarrow A_i)$ if and only if
  $\rho \models \beta$.

  (Only if). Assume that $\rho \in R'(\Sigma_{i=1}^n \ask(c_i)\rightarrow A_i)$.
  By Theorem~\ref{reqden},
  equation E4 in Table~\ref{densem} and
  Definition~\ref{defsemoperators} it follows that  $\rho\in D_1\cup D_2$
  where
  \[
  \begin{array}{lll}
  D_1 = & \{\rho \cdot \rho' \mid & \rho_j(O)=\rho_j(I)
  \mbox{ and } \rho_j(I)\not\vdash c_i
  \mbox{ for each } j\in [1,l(\rho)-1], i\in [1,n] \\
  &  & \rho_{l(\rho)}(I)\vdash c_h, \
  \rho_{l(\rho)}(O)=\rho_{l(\rho)}(I)
  \hbox { and } \rho'\in \os  D.A_h \cs'
  \hbox{ for an } h\in[1,n]\}  \mbox{ and }
  \\
  D_2 = & \{\rho \mid & \rho_j(O)=\rho_j(I) \mbox{ and }
  \rho_j(I)\not\vdash c_i
  \mbox{ for each } j\in [1,l(\rho)], i\in [1,n]\}
  \end{array}
  \]

  By definition of $\psi(A)$ and Definition~\ref{corrass}, if $\rho'\in \os  D.A_h \cs'$ then $\rho' \models \psi(A_h)$.
  Therefore, if $\rho \in D_1$ then $\rho \models
(\bigwedge_{j=1}^n \neg I_j \wedge stut)\; {\cal U}\; (I_i\wedge stut\wedge \bigcirc \psi (A_h))$ for an
$h\in[1,n]$ which implies $\rho \models \beta$ by definition of disjunction and of $\beta$.

If $\rho \in D_2$ then
  clearly $\rho \models\Box (\bigwedge_{j=1}^n\neg I(c_j)\wedge
  stut)$ and therefore $\rho \models \beta$. This complete the proof of the ``Only if'' part.
  The proof of the other implication is analogous (by using $D_1$ and $D_2$ above) and hence omitted.

  \item[{\bf (now $c$ then $A$ else $B$)}] By inductive hypothesis we obtain
  \[\Phi_1,\ldots,\Phi_n \vdash_p A\;sat\;\psi(A),
  \Phi_1,\ldots,\Phi_n \vdash_p B\;sat\;\psi(B) \ \mbox{ and }
  FV_{constr}(\psi(A))=FV_{constr}(\psi(B))=\emptyset.\]
  Therefore rule {\bf T3} implies that
  \[\Phi_1,\ldots,\Phi_n \vdash_p \mbox{\bf now $c$ then $A$ else $B$} \;sat\;
  (I(c)\wedge \psi(A))\vee (\neg I(c)\wedge \psi(B)).\]
  Since the inductive hypothesis implies also that
  $FV_{constr}((I(c)\wedge \psi(A))\vee (\neg I(c)\wedge
  \psi(B)))=\emptyset$, to prove the thesis we have only to show that
  \[(I(c)\wedge \psi(A))\vee (\neg I(c)\wedge \psi(B)) =
  \psi(\mbox{\bf now $c$ then $A$ else $B$})\] hold, that is,
  we have to show that $\rho \in R'(\mbox{\bf now $c$ then $A$ else $B$})$
  if and only if
  $\rho \models (I(c)\wedge \psi(A))\vee (\neg I(c)\wedge \psi(B))$.
  The proof is straightforward, by observing that from the definition
  of $\psi(A)$ and $\psi(B)$, from Definition~\ref{corrass},
  Theorem \ref{reqden}, equation {\bf E4} of Table~\ref{densem}
  and Definition~\ref{defsemoperators} it follows the following
  equality
  \[
  \begin{array}{lll}
   R'(\mbox{\bf now $c$ then $A$ else $B$})=& \{\rho \mid \rho_1(I)\vdash c  \mbox{ and }
   \rho \models \psi(A)\} \cup
   \\
   & \{\rho' \mid \rho'_1 (I)\not\vdash c \mbox{ and }
   \rho' \models\psi(B)\}.
  \end{array}
  \]

  \item[{\bf($\exists x \bf A$)}] The inductive hypothesis implies that
  \[\Phi_1,\ldots,\Phi_n \vdash_p A\;sat\;\psi(A)\ \mbox{ and }
  FV_{constr}(\psi(A))=\emptyset\]
 and therefore, by rule {\bf T4}, we obtain that
  \[\Phi_1,\ldots,\Phi_n \vdash_p A\;sat\;
  \exists x(\psi(A)\wedge {\it loc}(x))\wedge {\it inv}(x).\]
  From the inductive hypothesis and the definitions of
  ${\it loc}(x)$ and of ${\it inv}(x)$, we obtain that
  \begin{equation}\label{FVesiste}
  FV_{constr}(\exists x(\psi(A)\wedge {\it loc}(x))\wedge {\it inv}(x) )=\emptyset.
  \end{equation}
  In order to prove the thesis, we have then to show that
  \[\exists x(\psi(A)\wedge {\it loc}(x))\wedge {\it inv}(x) =
  \psi(\exists x A),\]
  holds, that is, we have to prove that $\rho \in R'(\exists x A)$ if and only if
  $\rho \models\exists x(\psi(A)\wedge {\it loc}(x))\wedge {\it inv}(x)$.

  Assume now that $\rho \in R'(\exists x A)$.
  Definition~\ref{corrass}, Theorem \ref{reqden},
  equation {\bf E6} of Table~\ref{densem},
  Definition~\ref{defsemoperators} and the definition of
  $\psi(A)$ imply that
  there exists $\rho'$ such that
  $\rho' \models\psi(A)$, $l(\rho)=l(\rho')$ and the
  following conditions hold:
  \begin{enumerate}
  \item for each $i\in [1,l(\rho)]$,
  $\exists x \rho_i(I)=\exists x \rho'_i(I)$ and $\exists x \rho_i(O)=\exists x
  \rho'_i(O)$;
  \item $\exists x \rho'_1(I)=\rho'_1(I)$ and for each $i\in [2,l(\rho)]$,
  $\exists x \rho_i'(I)\sqcup \rho'_{i-1}(O)=\rho'_i(I)$;
  \item for each $i\in [1,l(\rho)]$,
  $\rho_i(O) = \exists x \rho_i(O)\sqcup \rho_{i}(I)$.
 \end{enumerate}
 Now the proof is analogous to that one already given for the case of {\bf Rule T4} of
  Theorem \ref{exsoundness}.

  Conversely, assume that
  $\rho\models\exists x(\psi(A)\wedge {\it loc}(x))\wedge {\it inv}(x)$.
  Then the following facts hold:
\begin{enumerate}
  \item $\rho\models {\it inv}(x)$. Therefore, by definition of
  ${\it inv}(x)$, for each $i \in [1,l(\rho)]$ the following holds
  \begin{equation}\label{esiste1}
    \rho_i(O) = \exists x \rho_i(O)\sqcup \rho_{i}(I);
  \end{equation}
  \item $\rho\models\exists x(\psi(A)\wedge {\it loc}(x))$. Then, from case
  2 of Lemma~\ref{daungamma} and  (\ref{FVesiste}) we obtain that
  $\rho'\models\psi(A)\wedge {\it loc}(x)$ for some $\rho'$ such
  that
  \begin{equation}\label{esiste2}
    \exists x \rho'=\exists x \rho.
  \end{equation}
  Since $\rho'\models\psi(A)\wedge {\it loc}(x)$, from the
  definition of $\psi(A)$ and of
  ${\it loc}(x)$ it follows that
  \begin{eqnarray}
      &&\rho' \in R'(A) \ \mbox{and} \label{esiste3} \\
      &&\exists x \rho'_1(I)=\rho'_1(I) \mbox{ and for each }
      i\in [2,l(\rho)], \
      \exists x \rho_i'(I)\sqcup \rho'_{i-1}(O)=\rho'_i(I).
      \label{esiste4}
  \end{eqnarray}
  \end{enumerate}
  From Definition~\ref{corrass}, Theorem \ref{reqden}, equation {\bf E6} of
  Table~\ref{densem}, Definition~\ref{defsemoperators}, (\ref{esiste1}),
  (\ref{esiste2}), \ref{esiste3}) and (\ref{esiste4}), it follows that
  $\rho \in R'(\exists x A)$ and therefore the thesis holds.

\item[{\bf ($A_1\parallel A_2$)}] By inductive hypothesis we obtain that
  \[\Phi_1,\ldots,\Phi_n \vdash_p A_i\;sat\;\psi(A_i)
  \mbox{ and } FV_{constr}(\psi(A_i))=\emptyset, \]
  for $i=1,2$. Then by rule {\bf T5} we obtain also that
  \[\Phi_1,\ldots,\Phi_n \vdash_p A\;sat\; \exists{X,Y} (\psi(A_1)[X/O]\wedge
  \psi(A_2)[Y/O] \wedge par(X,Y)),\]
  where $X,Y \not \in FV(\psi(A_1))\cup FV(\psi(A_2))$ and $X\neq Y$.

  The inductive hypothesis and the definition of
  $par(X,Y)$ imply that
  \begin{equation}\label{FVparallelo}
  FV_{constr}(\exists{X,Y} (\psi(A_1)[X/O]\wedge
  \psi(A_2)[Y/O] \wedge par(X,Y))=\emptyset.
  \end{equation}
  Then, in order to prove the thesis we have to show that
  \[\exists{X,Y} (\psi(A_1)[X/O]\wedge \psi(A_2)[Y/O] \wedge par(X,Y))
  = \psi(A_1\parallel A_2),\] holds,
  that is, we have to prove that $\rho \in R'(A_1\parallel A_2)$ if and only if
  $\rho \models \exists{X,Y}(\psi(A_1)[X/O]\wedge \psi(A_2)[Y/O] \wedge par(X,Y))$.

  Assume that $\rho \in R'(A_1\parallel A_2)$.
  By Definition~\ref{corrass}, Theorem~\ref{reqden},
  equation E4 in Table~\ref{densem} and
  Definition~\ref{defsemoperators} it follows that there exist $\rho'\in R'(A_1)$ and
  $\rho'' \in R'(A_2)$ such that $l(\rho)=l(\rho')=l(\rho'')$ and,
  for each $i\in [1,l(\rho)]$, $\rho_i(I)=\rho'_i(I)=\rho''_i(I)$ and
  $\rho_i(O)=\rho'_i(O) \sqcup \rho'_i(O)$ hold.
  Since $\rho'\in R'(A_1)$, the definition
  of $\psi(A_1)$ implies that
  $\rho' \models \psi(A_1)$.
  Analogously we have that $\rho'' \models \psi(A_2)$.
  Now the proof is analogous to that one already shown for the case of  {\bf Rule T5}
  in the proof of Lemma~\ref{exsoundness}.

  Conversely, assume that $\rho \models\exists X,Y
  (\psi(A_1)[X/O] \wedge \psi(A_2)[Y/O] \wedge par(X,Y))$.
  We have to prove that $\rho \in R'(A_1\parallel A_2)$.
  By case 2 of Lemma~\ref{daungamma} and (\ref{FVparallelo})
  there exists a sequence of predicate assignments $\bar \rho$
  such that $\exists X,Y \bar \rho=\exists X,Y\rho$ and
  $\bar \rho\models \psi(A_1)[X/O]\wedge \psi(A_2)[Y/O] \wedge par(X,Y)$.

  We can now construct two new sequences $\rho'$ and $\rho''$ of predicate
  assignments having the same length as $\bar \rho$, such that, for each $i\in [1,l(\bar \rho)]$,
\begin{equation}\label{parallelo1}
  \rho'_i(O)=\bar \rho_i(X),  \ \ \rho''_i(O)=\bar \rho_i(Y)
  \mbox{ and }
  \rho'_i(Z)=\rho''_i(Z)= \bar \rho_i(Z),
  \mbox{ for each $Z \in M$, $Z \neq O$.}
\end{equation}
  Since $X,Y \not \in FV(\psi(A_1))\cup FV(\psi(A_2))$ and $X\neq
  Y$, by construction it follows that $\rho'\models \psi(A_1)$ and
  $\rho''\models \psi(A_2)$. Therefore, by definition of $\psi(A)$,
\begin{equation}\label{parallelo2}
  \rho'\in R'(A_1) \mbox{ and } \rho''\in R'(A_2).
\end{equation}
  Moreover, since
  $\bar \rho\models par(X,Y)$, again by construction we obtain that, for $i\in [1,l(\bar
  \rho)]$,
\begin{equation}\label{parallelo3}
  \bar \rho_i(O)=\rho'_i(O) \sqcup \rho''_i(O)
\end{equation}
  holds. From Definition~\ref{corrass}, Theorem~\ref{reqden}, equation E4 in
  Table~\ref{densem}, Definition~\ref{defsemoperators},
  (\ref{parallelo1}), (\ref{parallelo2}) and (\ref{parallelo3}) it follows that
  $ \bar \rho \in  R'(A_1\parallel A_2)$.
Observe now that, by definition, for any sequence $\rho$ of
predicate assignments and for any tccp process $A$, $\rho \in
R'(A)$  if and only if its `projection' on the distinguished
predicates $I$ and $O$
  generates a reactive sequence of $R(A)$.
  Hence, since $\exists X,Y \bar \rho=\exists X,Y\rho$ and
  $ \bar \rho \in  R'(A_1\parallel A_2)$, we have that
  $\rho \in  R'(A_1\parallel A_2)$ and therefore the thesis holds.

\item[{\bf ($A= p(y)$)}] Immediate.
\end{description}

\bigskip

From the above Theorem we derive the following corollary:

\begin{corollary}[Completeness]\label{cor:completeness}
Let $D=\{p_1(x_1)\;::\;A_1,\ldots , p_n(x_n)\;::\;A_n\}$ and
$A_{n+1}$ be an agent such that $A_i$, for $1\leq i\leq n+1$, only
involves calls to procedures declared  in $D$. It follows that
$\models D.A_{n+1}\;sat\;\phi$ implies  $\vdash_p D.A_{n+1}
\;sat\;\phi$.
\end{corollary}
{\bf Proof}
 By the above lemma we have for $i=1,\ldots,n$ that
\[
\Phi_1,\ldots,\Phi_n \vdash_p A_i\;sat\;{\it \psi}(D.p_i(x_i))
\]
where $\Phi_i = p_i(x_i)\;sat\;{\it \psi}(D.p_i(x_i))$ (note that
${\it \psi}(D.A_i)={\it \psi}(D.p_i(x_i))$). Now a repeated
application of the recursion rule gives us $\vdash_p
D.p_i(x_i)\;sat\;{\it \psi}(D.p_i(x_i))$. Again by the above lemma
we have that
\[
\Phi_1,\ldots,\Phi_n \vdash_p A_{n+1}\;sat\;{\it \psi}(D.A_{n+1})
\]
It follows  by a straightforward induction
on the length of the derivation (which does not involve
applications of the recursion rule) that
\[
\Phi'_1,\ldots,\Phi'_n \vdash_p D.A_{n+1}\;sat\;{\it
\psi}(D.A_{n+1})
\]
where $\Phi'_i=D.  p_i(x_i)\;sat\;{\it \psi}(D.p_i(x_i))$, $i=1,\ldots,n$. So, since $\vdash_p \Phi'_i$,
$i=1,\ldots,n$, we thus derive that $\vdash_p D.A_{n+1}\;sat\;{\it \psi}(D.A_{n+1})$. Assume now that $\models
D.A_{n+1}\;sat\; \phi$. Then $\models {\it \psi}(D.A_{n+1})\rightarrow\phi$ holds and an application of {\bf C7}
gives us $\vdash_p D.A_{n+1}\;sat\;\phi$.
\\

\bigskip

The formula $\psi(A)$ (analogous to the strongest postcondition)
has been used to prove completeness. However, often to prove a
property of a program it is sufficient to deal with some simpler
property. The situation can be compared to the problem of
finding the suitable invariant when using the standard Hoare
systems for imperative programming.

\section{Related and future work}

We introduced a temporal logic for reasoning about the correctness of a timed extension of ccp and we proved the
soundness and completeness of a related proof system.

A simpler temporal logic for tccp has been defined in \cite{BGM01time} by considering epistemic operators of
``belief'' and ``knowledge'' which corresponds to the operators $I$ and $O$ considered in the present paper. Even
though the intuitive ideas of the two papers are similar, the technical treatment is different. In fact, the logic
in \cite{BGM01time} is less expressive than the present one, since it does not allow constraint (predicate)
variables. As a consequence, the proof system defined in \cite{BGM01time} was not complete.

Recently, a logic for a different timed extension of ccp, called
ntcc, has been presented in \cite{PV01}. The language ntcc
\cite{Va00,NV02} is a non deterministic extension of the timed
ccp language defined in \cite{SJG96}. Its computational model,
and therefore the underlying logic, are rather different from
those that we considered. Analogously to the case of the ESTEREL
language, computation in ntcc (and in the language defined in
\cite{SJG96}) proceeds in ``bursts of activity'': in each phase
a ccp process is executed to produce a response to an input
provided by the environment. The process accumulates
monotonically information in the store, according to the
standard ccp computational model, until it reaches a ``resting
point'', i.e. a terminal state in which no more information can
be generated. When the resting point is reached, the absence of
events can be checked and it can trigger actions in the next
time interval. Thus, each time interval is identified with the
time needed for a ccp process to terminate a computation.
Clearly, in order to ensure that the next time instant is
reached, the ccp process has to be always terminating, thus it is
assumed that it does not contain recursion (a restricted form of
recursion is allowed only across time boundaries). Furthermore,
the programmer has to transfer explicitly the all information
from a time instant to the next one by using special primitives,
since at the end of a time interval all the constraints
accumulated and all the processes suspended are discarded,
unless they are argument to a specific primitive. These
assumptions allow to obtain an elegant semantic model consisting
of sequences of sets of resting points (each set describing the
behavior at a time instant).

On the other hand, the tccp language that we consider has a
different notion of time, since each time-unit is identified with
the time needed for the underlying constraint system to
accumulate the tell's and to answer the ask's issued at each
computation step by the processes of the system. This assumption
allows us to obtain a direct timed extension of ccp which
maintain the essential features of ccp computations. No
restriction on recursion is needed to ensure that the next time
instant is reached, since at each time instant there are only a
finite number of parallel agents which can perform a finite
number of (ask and tell) actions. Also, no explicit transfer of
information across time boundaries is needed in {\it tccp},
since the (monotonic) evolution of the store is the same as in
ccp (these differences affects the expressive power of the
language, see \cite{BGM00} for a detailed discussion). Since the
store grows monotonically, some syntactic restrictions are
needed also in tccp in order to obtain bounded response time,
that is, to be able to statically determine the maximal length
of each time-unit (see \cite{BGM00}).

>From a logical point of view, as shown in \cite{BGMP94} the set of
resting points of a ccp process characterizes essentially the
strongest post condition of the program (the characterization
however is exact only for a certain class of programs). In
\cite{PV01} this logical view is integrated with (linear) temporal
logic constructs which are interpreted in terms of sequences of
sets of resting points, thus taking into account the temporal
evolution of the system. A proof system for proving the resulting
linear temporal properties is also defined in \cite{PV01}. Since
the resting points provide a compositional model (describing the
final results of computations), in this approach there is no need
for a semantic and logical representation of ``assumptions''. On
the other hand such a need arises when one wants to describe the
input/output behavior of a process, which for generic (non
deterministic) processes cannot be obtained from the resting
points. Since tccp maintains essentially the ccp computational
model, at each time instant rather than a set of final results
(i.e. a set of resting points) we have an input/ouput behavior
corresponding to the interaction of the environment, which
provides the input, with the process, which produces the output.
This is reflected in the the logic we have defined.

Related to the present paper is also \cite{FPV00}, where tcc
specifications  are represented in terms of graph structures in
order to apply model checking techniques. A finite interval of
time (introduced by the user) is considered in order to obtain a
finite behavior of the tcc program, thus allowing the
application of existing model checking algorithms.

Future work concerns the investigation of an axiomatization for 
the temporal logic
introduced in this paper and the possibility of obtaining decision procedures, for example considering  a semantic
tableaux method. Since {\em reactive sequences} have been used also in the semantics of several other languages,
including dataflow and imperative ones \cite{Jo85,BP90c,BKPR91a,Br93}, we plan also to consider extensions of our
logic to deal with these different languages.

\end{document}